\documentclass[11pt,prd,superscriptaddress,nofootinbib]{revtex4}
\usepackage{graphicx}
\usepackage[usenames,dvipsnames]{xcolor}
\begin{document}

	\begin{center}
		{\Large \bf Axial-vector form factor of nucleons at finite temperature from the  AdS/QCD soft-wall model}
		\vskip 0.5 cm
				
		{Shahin Mamedov$^{a,b,c}$ \footnote{e-mail : sh.mamedov62@gmail.com (corresponding author)} and
			Narmin Nasibova$^b$ \footnote{e-mail :n.nesibli88@gmail.com}}
		\vskip 0.5cm
		
		{ \it \indent $^a$ Institute for Physical Problems, Baku State University, \\ Z. Khalilov str. 23, Baku, AZ1148, Azerbaijan}\\
		\it \indent $^b$ Institute of Physics, Ministry of Science and Education,\\
		AZ1143, H. Javid Avenue 131,  Baku, Azerbaijan\\
  \it \indent $^c$ Department of Physics and Electronics, Khazar Universty,\\
		41, Mehseti Street,  Baku, AZ-1096, Azerbaijan\\.
	\end{center}
	
	\thispagestyle{empty}
	
	\vskip2cm
	
	\centerline{\bf Abstract} 
	\vskip 4mm
The axial-vector form factor of the nucleons is studied at a finite temperature using the holographic soft-wall model with the thermal dilaton field. We use the bulk interaction action known from the zero temperature case and apply the profile functions of fields thermalized by the interaction with the thermal dilaton field. The dependence of the axial form factor on the square of the transferred momentum and the temperature is plotted for the ground and excited states of the nucleons.
	\vspace{1cm}
	\section{Introduction}
	In the weak processes, such as $\beta$ decay  $n \rightarrow p + e^{-}+ \tilde{\nu}_{e}$ or $\mu$ capture $\mu^{-}+p=\nu_{\mu}+n$, an interaction takes place with the axial-vector current of the baryons. The form factor $G_A\left(Q^2\right)$, which describes the interaction vertex in these processes, is called the axial-vector form factor of the nucleons. The axial-vector form factor is also used for the interpretation of neutrino experiments \cite {Mosel, Alvarez}. The form factor has been studied in the framework of different models of field theories  (QCD sum rules, chiral quark model, etc.), for the low or high values of the momentum transfer $Q^2$ in the works \cite{Bernard, Alexandrou, Eichmann, Anikin, Meyer, Ramalhoa, Chen, H. Hashamipour}. After the appearance of the QCD models based on AdS/CFT correspondence \cite{Maldacena, Gubser, Witten}, which are called as holographic QCD, or AdS/QCD models \cite{Boschi1, Boschi2, Erlich, Rold1, Rold2, Karch, Brodsky,{Branz}}, many problems of strong interactions \cite{Paula, Rinaldi} have been considered in the framework of these models. The holographic QCD enables us to investigate the axial-vector form factor by imposing no limitation on $Q^2$, i.e., to consider it at both small and large values of the momentum transfer. As a result, both momentum areas were joined within one graph, which nicely agrees with the graphs of the different approaches for the various  $Q^2$ areas. It explains the experimental data as well. Within holographic QCD the $G_A$ form factor was considered in the hard-wall model framework in Refs. \cite{Mamedov, Gutsche1, Atayev1,{Huseynov}} and in the soft-wall model framework in Refs. \cite{Atayev1, Lyubovitskij}. The main advantage of the holographic soft-wall model is that it produces linear Regge trajectories in the mass spectrum of the mesons \cite{Karch, Paula, Rinaldi}. In Ref. \cite{Atayev2} the axial form factor splitting in the holographic approach has been studied in the so-called isospin medium, which is a simplified model of the nuclear medium. Such a "medium" takes into account only the isospin chemical potential of the medium and neglects all other quantities. Interaction with the isospin background of the medium leads to a mass splitting of nucleons, which results in the splitting of the $G_A$ form factor into two branches, each of which corresponds to a certain isospin projection.
    Recently, in Refs. \cite{Gutsche4, Nasibova1}, the finite temperature theory of mesons and baryons was considered within the soft-wall model of the AdS/QCD. Explicit formulas, obtained there for the wave functions of mesons and baryons in 5D, enable us to solve 4D QCD interaction issues at finite temperatures up to critical temperature $T_{c}=0.2$ $GeV$. It is well known that, the critical temperature describes the confinement-deconfinement phase transition in quantum chromodynamics. In the dual AdS gravity side, this temperature corresponds to the temperature of a first-order Hawking-Page phase transition between the thermal AdS space-time and black hole. The critical temperature has been estimated as $T_{c}=0.122$ in the hard wall model, $T_{c}=0.191$  $GeV$ in soft wall model  of holographic QCD in Ref. \cite{Gutsche2}, $T_{c}=0.192\pm0.007\pm0.004$ $GeV$ in the lattice QCD \cite {Cheng5}.
    
   The study of the medium modification of nucleon properties  is a longstanding topic \cite{Rakhimov}.  With the application of the holographic QCD approach  the hadron, electromagnetic, and gravitational form factors of the nucleons were studied in Refs. \cite{Gutsche2,Gutsche3,Gutsche4,Allahverdiyev,{Gutsche5}, {Abidin}}. Moreover, the temperature dependencies of meson-nucleon coupling constants were found in Refs. \cite{Nasibova1,Taghiyeva,Nasibova}.

   This approach enables us to study the $G_A$ form factor in the thermal medium. The same problem was considered in Ref. \cite{Dominguez} within the QCD sum rules framework. Here we aim to consider a temperature dependence of the axial-vector form factor neglecting all other medium quantities (density, chemical potential, e.t.c.); however, present study will be useful for the neutrino- and anti-neutrino-nucleus scattering experiments Refs. \cite{Gysbers,Lovato,King}.

The remainder of this paper is organized as follows:

In Section II, we briefly present the definition of the axial-vector form factor at finite temperature. 
Sec. III is devoted to the basic notions in the holographic soft-wall model with the thermal dilaton field. 
Sec. IV concerns  the profile functions of fermion fields at finite temperatures. In Sec. V,  the breaking of chiral symmetry is described at a finite temperature. In Sec. VI, we obtain the equation of motion of meson bulk-to-boundary propagator at finite temperature. In Sec. VII, we give an interaction Lagrangian between the bulk fields. Using the AdS/CFT correspondence, we obtain an integral expression for the form factor. In Sec. VIII, we derive the $G_A(T)$ form factor at a finite temperature from the corresponding action.
In Sec. IX, the free parameters are fixed, and the axial-vector form factors' normalized graphs are plotted for the different values of the parameter $\alpha$, which serves as a  parameter of the small perturbative correction to the quadratic dilaton \cite{Gutsche2, Gutsche3}.
In the last section, we summarize our results.
	\section{Axial-vector current of the nucleons}
	In QCD the isovector axial-vector current of nucleons  is defined as follows \cite{Bernard}:
	\begin{equation}
		j^{\mu ,a}\left( x\right) =\overline{\psi }\left( x\right) \gamma ^{\mu
		}\gamma^{5}\frac{\tau ^{a}}{2}\psi \left( x\right), 
		\label{1}
	\end{equation}
	where  $\psi $ is the doublet of $u$ and $d$ quarks $\psi =\left(
	\begin{array}{c}
		u \\
		d
	\end{array}
	\right)$ and $\tau^a$ are the isospin Pauli matrices. The current (\ref{1}) is a partially conserved:
	\begin{equation}
		\partial _{\mu }j^{\mu ,a}\left( x\right) =i\overline{\psi }\left( x\right)
		\gamma^{5}\left\{ \frac{\tau^{a}}{2},\mu \right\} \psi \left( x\right).\label{2}
	\end{equation}
	Here $\mu $ is the mass matrix of the up and down quarks, which has the diagonal form: $\mu =diag\ \left( m_{u},m_{d}\right) $. In the exact isospin symmetry case, these masses are equal $\left( m_{u}=m_{d}=m\right) $ and the $\mu$ matrix is proportional to the unit matrix. A matrix element of the nucleon isovector axial-vector current at finite temperature can be written analogous to the zero temperature case \cite {Bernard}:
	\begin{equation}
		\left\langle N\left( p^{\prime },T\right) \left| j^{\mu ,a}\left(
		0\right) \right| N\left( p,T\right) \right\rangle =\overline{u}\left(
		p^{\prime }\right) \left[ \gamma ^{\mu }\gamma^{5}G_{A}\left( q^{2}, T\right) +\frac{q^{\mu }}{2m_{N}}\gamma^{5}G_{P}\left( q^{2}, T\right) \right] \frac{\tau ^{a}}{2}u\left( p\right).  \label{3}
	\end{equation}%
	Here $q_{\mu }=p_{\mu }^{\prime }-p_{\mu }$ is the total momentum in the interaction vertex and $m_{N}$ is the nucleon mass, $G_{A}\left( q^{2}, T\right) $ and $G_{P}\left(q^{2}, T\right) $ are called the axial-vector and induced pseudoscalar form factors respectively. Due to the Hermiticity of the $j^{\mu ,a}$ current the form factors $\ G_{A}\left( q^{2}, T\right) $ and $G_{P}\left(q^{2}, T\right) $ are real functions of $q^{2}$ in the $q^{2}\leq 0$ domain. We shall consider the  $Q^2=-q^2$ timelike momentum domain. Here we consider the form factor $\ G_{A}\left( q^{2}, T\right) $ at finite temperature in the soft-wall AdS/QCD model framework. 
	\section{Soft-wall model at finite temperature}
	The action for the soft-wall model, which we use for the present finite-temperature study, contains a dilaton field $\varphi$ depending on the temperature:
	\begin{equation}
		S=\int d^{4}x dz\sqrt{g}e^{-\varphi(z,T)}L(x,z,T),\label{4}
	\end{equation}
	where  $g=|det g_{M N}|$, $(for M,N=0,1,2,3,5)$  and the extra dimension $z$ varies in the range $0 \leq z <   \infty $.
	The background geometry is assumed to be 5D AdS space-time with the metric at finite temperature:  
	\begin{equation}
		ds^{2}=e^{2A(z)}\left[f(z,T)dt^{2}-d{\vec{x}}^{2}-\frac{dz^{2}}{f(z,T)}\right].\label{5}
	\end{equation}
	Here 
	$ x=(t,\vec{x})$ is the set of Minkowski coordinates, $z$  is the holographic coordinate,
	$A(z)=log(R/z)$, and  $R$ is the AdS radius.
	The thermal factor $f(z,T)$ has a form \cite {Witten2}:
	\begin{equation}
		f(z,T)=1-\frac{z^{4}}{z_{H}^{4}}, \label{6}
	\end{equation}
	where $z_{H}$ is the position of the event horizon and is related to the Hawking temperature as $T=1/(\pi z_{H})$. The dilaton field $\varphi(z)=k^2z^2$, where $k$ is a scale parameter of a few hundred $MeV$, and introduced to make the integral over the $z$ finite at IR boundary $(z\rightarrow \infty )$. For the finite-temperature soft-wall model, the form of the dilaton has been modified by including the temperature-depending terms. In this model
	it is convenient to apply the Regge-Wheeler tortoise coordinate $r$,
	\begin{equation}
		r=\int \frac{dz}{f(z)}\label{7}
	\end{equation}
	instead of $z$ \cite{Regge}, and  to neglect the  terms of higher order than $T^{8}$  in the expansion of $z$. This gives following relation between the  $r$ and  $z$ coordinates \cite {Gutsche2}:
	\begin{equation}
		r\approx z\left[1+\frac{z^{4}}{5z_{H}^{4}}+\frac{z^{8}}{9z_{H}^{8}}\right].\label{8}
	\end{equation}
	The metric for the AdS-Schwarzschild space-time in these coordinates will be written in the form: 
	\begin{equation}
		ds^{2}=e^{2A(r)}f^{\frac{3}{5}}(r)\left[dt^{2}-\frac{\left(d\vec{x}\right)^{2}}{f(r)}-dr^{2}\right] \label{9}
	\end{equation}
	with $A(r)=log(R/r)$. The thermal factor $f(r)$ depending on the $r$ coordinate has same form as in Eq. (\ref{7}):
	\begin{equation}
		f(r)=1-\frac{r^{4}}{r_{H}^{4}}.
		\label{10}
	\end{equation}
	
	The AdS-Schwarzchild geometry is more suitable for high
$T$, while for low temperatures this metric can be also used to generate a
small $T$-expansion. The limit $T = 0$ corresponds to a mapping
of the AdS-Schwarzchild geometry onto AdS Poincare metric,
and the small $T$-the behavior of hadron properties can be generated
in the formalism based on the AdS Poincare metric, and with
the use of a thermal dilaton. It can be shown that the results will coincide since it leads to equivalent results: AdS Schwarzchild
geometry with small $T$ is equal to AdS Poincare
metric with thermal dilaton.
	
	The thermal version of the 	usual quadratic dilaton was applied in Ref. \cite {Vega}
	\begin{equation}
		\varphi(r,T)=K^{2}(T)r^{2}
		\label{11}
	\end{equation}
	and  developed further in Refs. \cite{Gutsche2,Gutsche3,Gutsche4,Nasibova}. As expected, by setting $T = 0$  the terminal dilaton can be reduced to the usual one $\varphi(r,0)=k^{2}r^{2}$.
	The $K^{2}(T)$ parameter in Eq. (\ref{11}) is the parameter of spontaneous breaking of chiral symmetry whose explicit form was established in Refs. \cite{Gutsche2,Gutsche3,Gutsche4,Nasibova}:
	\begin{equation}
		K^{2}(T)=k^{2}[1+\rho(T)+O(T^{6})].
		\label{12}
	\end{equation}
In section VI, we shall bring the $K^{2}(T)$ dependence in detail. The $\rho(T)$ function in Eq.(\ref{12}) encodes the $\emph T$ dependence of the dilaton field and was found in the form:
	\begin{equation} 
		\rho(T)=\frac{9\alpha\pi^{2}}{16}\frac{T^{2}}{12F^{2}}-\frac{ N_{f}^{2}-1}{N_{f}}\frac{T^{2}}{12F^{2}}-\frac{N_{f}^{2}-1}{2N_{f}^{2}}\left(\frac{T^2}{12F^2}\right)^{2} +O\left(T^6\right).
		\label{13}
	\end{equation}
	Here  $N_{f}=2$ is the number of quark flavors. The pion decay constant  $F$ is  proportional to the $k$  parameter $F=\frac{k\sqrt{6}}{8}$ \cite{Gasser}. The relations (\ref{12})-(\ref{13})  were obtained by introducing the thermal prefactor $ e^{-\lambda_{T}}$ with  
 $$\lambda_{T}(z)=\alpha\frac{z^{2}}{z^{2}_{H}}+\gamma\frac{z^{4}}{z^{4}_{H}}+\xi\frac{k^{2}z^{6}}{z^{4}_{H}}$$ 
  in Ref. \cite{Gutsche2}. The parameter $\gamma$ was fixed to guarantee the gauge invariance and massless ground states pseudoscalar mesons in chiral limit and the parameter $\xi$ was fixed to drop the radial dependence in six power. Thus, the thermal dilaton depends only on parameter $\alpha$. This small parameter encodes the contribution of gravity to the restoration of chiral symmetry at the critical temperature $T_c=0.2$ $GeV$.
	\section{Nucleons at finite temperature}
	In this section, we briefly present the profile function for a fermion field in the model with thermal dilaton, which was derived in Ref. \cite {Gutsche3}. Action for the bulk fermion field in this model is written in the form:
	\begin{equation}
		S=\int {d^{4}x}dre^{-\varphi(r, T)}\sqrt{g}{\bar{\Psi}}(x,r,T)D_{\pm }(r)\Psi(x,r,T).
		\label{16}
	\end{equation}
	Here $D_{\pm}(r) =\frac{i}{2}\Gamma^{M}[\partial_{M}-\frac{1}{4}\omega_{M}^{ab}\ [\Gamma_{a}\Gamma_{b}]]\mp[\mu(r, T)+U_{F}(r, T)]$ are covariant derivatives, which include five-dimensional temperature-dependent mass $\mu (r, T)=\mu\ f^{\frac{3}{10}}(r, T)$.  $\mu=\emph N_{B}+\emph L-\frac{3}{2}$, where $\emph N_{B}=3$  and $\emph L$ are the number of partons and orbital angular momentum $(\emph L=0)$ of the nucleons, respectively.
	Thermal potential for nucleons $U_{F}(r, T)$ is defined as $U_{F}(r, T)=\varphi(r, T)/f^{\frac{3}{10}}(r, T)$. Non-zero components of the spin connection $ \omega_{M}^{ab} $ are given by $
	\omega_{M}^{ab}=(\delta_{\mu }^{a}\delta_{r}^{b}-\delta_{\mu }^{b}\delta_{r}^{a})\ r f^{\frac{1}{5}}(r, T)$.
	Commutator of the Dirac matrices is defined by $\sigma^{MN}=[\Gamma^{M},\Gamma^N]$ and $\Gamma^{M} =e_{a}^{M}\Gamma^{a}$, where
	$e_{a}^{M}=\ z\delta_{a}^{M}$ is the inverse vielbein. The $\Gamma^{a}$ matrices are defined as $ \Gamma^{a}=(\Gamma^{\mu },\ -i\Gamma ^{5}) $.    
	The $\Psi$ field is decomposed into the left and right chirality components as $\Psi(x,r, T)=\Psi^{R}(x,r, T)+\Psi^{L}(x,r, T)$, with definition $\Psi^{L, R}(x,r, T)= \frac{1 \mp \gamma^{5}}{2}\Psi$,  where
	$\gamma^{5}\Psi^{L}=-\Psi^{L}$, $\gamma^{5}\Psi^{R}=\Psi^{R}$. 
	Kaluza-Klein decomposition of the $\Psi^{L,R}$ fields is written as $\Psi^{L,R}=\sum_{n}	\Phi_{n}^{L/ R}(r,T)\psi_n(x)$, where $\Phi_{n}^{L/ R}(r,T)$ are the profile functions and 4D wave functions $\psi_n(x)$ satisfy free Dirac equation $\not{p} \psi_n(x)= M_{n}(0)\psi_n(x)$.  In further calculations, the following replacement in the profile functions is useful \cite {Gutsche3}:
	\begin{equation}
		\Phi_{n}^{L/ R}(r,T)=e^{-\frac{3}{2}A(r)}F_{n}^{L/ R}(r,T).\label{17}
	\end{equation}
	Equation of motion (EOM) obtained from the action in Eq. (\ref{16}) is the 5D Dirac equation at finite temperature. Substituting $\Phi_{n}^{L/ R}(r,T)$ in EOM following equation may be obtained for the  $F_{n}^{L/R}(r,T)$ profiles in the nucleon's rest frame \cite {Gutsche3}:
	\begin{equation}
		\left[\partial_{r}^2+U_{L/R}(r,T)\right]F_{n}^{L/R}(r,T)=M_{n}^{2}(T)F_{n}^{L/R}(r,T). \label{18}
	\end{equation}
Here $U_{L/R}(r, T)$ are the effective potentials written as the sum of the zero- and finite-temperature potential terms:
	\begin{equation}
		U_{L/R}(r, T) =U_{L/R}(r)+\Delta U_{L/R}(r,T)  \label{19}
	\end{equation}
	with the explicit forms
	\begin{equation}
		U(r)=k^{4}r^{2}+\frac{(4m^{2}-1)}{4r^{2}},
		\ \Delta U(r,T)=2\rho(T)k^{4}r^{2}.
		\label{20}
	\end{equation}
	Here $m=N+L-2$. Particularly, for the nucleon with three partons  $m=L+1$.
	$ M_{n}^{2}$ gives the Eq. (\ref{18}) is the nucleon  mass spectrum, and at the low temperatures can also be written as the sum of the zero- and finite-temperature parts:
	\begin{equation}
		M_{n}^{2}(T) =\ M_{n}^{2}(0)+\Delta M_{n}^{2}(T),
		\label{21}
	\end{equation}
	where
	\begin{equation}
		M_{n}^{2}(0)=4k^2\left(n+\frac{m+1}{2}\right),
		\label{22}
	\end{equation}
	\begin{equation}
		\Delta M_{n}^{2}(T)=\rho(T)M_{n}^{2}(0) + \frac{R\pi^{4}T^{4}}{k^{2}} 
		\label{23}
	\end{equation}
 $R =(6n-1)(m+1)$.
	Equation (\ref{18}) can be solved by using the boundary conditions on $F_{n}^{L/R}(r, T)$ in the ultraviolet (UV) and infrared  (IR)  limits
	at small $r$
	\begin{equation}
		F_{n}^{L/R}(r,T)\sim r^{N+L-1\pm\frac{1}{2}},
		\label{24}
	\end{equation}
	and at large $r$ as
	\begin{equation}
		F_{n}^{L/R}(r,T)\rightarrow0.
		\label{25}
	\end{equation}
	The normalization conditions for the profile functions $F_{n}^{L/R}(r,T)$ are:
	\begin{equation}
		\int_{0}^{\infty } dr e^{-3A(r)}\Phi_{m}^{L,R}(r,T)\Phi_{n}^{L,R}(r,T))=\int_{0}^{\infty } dr F_{m}^{L,R}(r,T) F_{n}^{L,R}(r,T)=\delta_{mn}.
		\label{26}
	\end{equation}
	In general the Schrodinger-type equations  (\ref{18})  have following analytical solutions:
	\begin{equation}
		F_{n}^{L/R}(r,T)=\sqrt{\frac{2\Gamma (n+1)}{\Gamma (n+m_{L/R}+1)}}K^{m_{L/R}+1}r^{m_{L/R}(T)+\frac{1}{2}}e^{-\frac{K^{2}r^{2}}{2}}L_{n}^{m_{L/R}}\left(K^{2}r^{2}\right). 
		\label{27}
	\end{equation}  
	By taking here $m_{L/R}=m\pm \frac{1}{2}$ in it, we obtain an expression for the profile functions of the nucleons, which coincide with $T\rightarrow0$ limit with the ones obtained in Ref. \cite{Gutsche11} for the $T=0$ case.
	
	\section{Condensate, dilaton and chiral symmetry breaking}
	The action for the pseudo-scalar $X$ field in AdS/QCD has a form:
	\begin{equation}
		S=\int_{0}^{\infty}d^{5}x\sqrt{g}e^{-\varphi(z)}Tr\{|DX|^{2}-m_{5}^{2}|X|^{2}\}.
		\label{28}
	\end{equation}
	In terms of the tortoise coordinate $r$ and with the thermal dilaton this action will slightly change:
	\begin{equation}
		S_{X}=\int d^{4}xdr\sqrt{g}e^{-\varphi(r,T)}Tr\left[|DX|^{2}+3|X|^{2}\right],
		\label{29}
	\end{equation}
	where $DX$ is the covariant derivative including the interaction with the gauge fields $A_{L, R}$. In terms of  the vector $M_M$ and  the axial-vector $A_M$ fields the covariant derivative is defined as: $D^{M}X=\partial^{M}X-iA_{L}^{M}X+iXA_{R}^{M}=\partial^{M}X-i\left[M_M,X\right]-i\{A_M,X\}$, $A_{L,R}^{M}=A_{L,R}^{M}t^{a}$ and $F_{L,R}^{MN}$ are the field strengths of these fields. The $X$ field transforms under the bifundamental representation of the flavor symmetry group $SU(2)_{L}\times SU(2)_{R}$ of the model and performs the breaking of the chiral symmetry by Higgs mechanism \cite{Gherghetta}. EOM for the $X$ field, which is obtained from Eq. (\ref{28}), has a solution 
	\begin{equation}
		\langle X \rangle=\frac{1}{2}v(z).
		\label{30}
	\end{equation}
	In the soft-wall model, for $v(z)$ in $z\rightarrow 0$ limit different authors \cite{Karch, Zhang, Gutsche30, Colangelo} apply the solution, which coincides with the one found within the hard-wall model:
	\begin{equation}
		v(z)=\frac{1}{2}M_{q}az+\frac{1}{2a}\Sigma z^{3}.  
		\label{31}
	\end{equation}
	Here $a=\sqrt{N_c}/(2\pi)$  $(N_c=3)$ is the normalization parameter \cite{Cherman}. According to the dictionary of the bulk/boundary correspondence, the parameters $M_{q}$ and  $\Sigma$ are identified with the $u,d$ quark mass matrix and with the chiral condensate $\Sigma=<0|\bar{q}q|0> $, respectively.
	For the finite-temperature case the quark condensate within $v(z)$ in Eq. (\ref{31}) depends on temperature; whereas the $z$ coordinate should be replaced by $r$. Then, the $v(r,T)$ solution has a form: 
	\begin{equation}
		v(r,T)=\frac{1}{2}M_{q}ar+\frac{1}{2a}\Sigma(T) r^{3}. 
		\label{32}
	\end{equation}
	To the function $\Sigma(T)$, it was found in Ref. \cite{Gasser} by using two-loop chiral perturbation theory at finite temperature, and applied in the Refs. \cite{Gutsche2,Gutsche3} for the finite-temperature soft-wall model. It has a form:
	\begin{equation}
		\Sigma(T)=\Sigma[1-\frac{N_{f}^{2}-1}{N_{f}}\frac{T^{2}}{12F^{2}}-\frac{N_{f}^{2}-1}{2N_{f}^{2}}(\frac{T^{2}}{12F^{2}})^{2}+O(T^{6})]=\Sigma[1+\Delta_{T}+O(T^{6})], 
		\label{33}
	\end{equation}
	where $N_{f}$ is the number of quark flavors and $F$ is the pion decay constant. As is seen from Eq. (\ref{31}) $\Sigma(T=0)=\Sigma$. It can be seen that
$$\Delta_{T}=-\frac{N_{f}^{2}-1}{N_{f}}\frac{T^{2}}{12F^{2}}-\frac{N_{f}^{2}-1}{2N_{f}^{2}}(\frac{T^{2}}{12F^{2}})^{2}$$
 in Eq. (\ref{33}) is related to the $\rho(T)$ in Eq. (\ref{13}).
	In the soft-wall model the dilaton field  $\varphi(r)$ is responsible for the dynamical breaking of the chiral symmetry. The chiral quark condensate $\Sigma(T)$ is the result of chiral symmetry breaking. Both constants $K$ and $\Sigma$ depend only on $T$. Therefore, it was supposed in Refs. \cite{Gutsche2,Gutsche3} that the $T$-dependence of the dilaton parameter  $K^2(T)$  should be similar to $\Sigma(T)$:
	\begin{equation}
		K^2 (T)=k^2\frac{\Sigma(T)}{\Sigma}.
		\label{34}
	\end{equation}
 It was also conjectured that the relation 
	
	\begin{equation}
		\Sigma=-N_{f}BF^{2}
		\label{35}
	\end{equation}
written for the $T=0$ case holds for the $T\neq0$ case also:
	\begin{equation}
		\Sigma(T)= - N_{f}B(T)F^{2}(T).
		\label{36}
	\end{equation}
	Here $B$ is the condensate parameter. Then, as a result of Eqs. (\ref{11}) and (\ref{34}) the $T$-dependence of the $\Sigma(T)$ condensate can be expressed in terms of the $\Delta(T)$ function \cite{Gherghetta}:
	\begin{equation}
		\Sigma(T)=\Sigma\left[1+\Delta(T)\right]+O(T^{6}).
		\label{37}
	\end{equation}
	 The temperature dependence of $ F(T)$ and  $B(T)$ relation have been studied in Ref. \cite{Gutsche2}.
 \section{Meson propagator at finite temperature}
	It is known from the zero-temperature case, the EOM for the $A_M$ axial-vector field in the soft-wall model coincides with the one for the $V_M$ vector field at certain limits $(m_q\rightarrow 0, z\rightarrow z_{UV})$ \cite{Rold2, Karch, Taghiyeva}. The Yukawa interaction term, which is proportional to $v(z)$, is usually neglected in both hard-wall \cite{Maru} and soft-wall \cite{Gutsche2, Gutsche3} models when solving the EOM for the fermion field. Here we accept the approximation where the function $v(z)$ is not taken into account to get solutions of the EOM for the axial vector field as well.  Of course, the bulk-to-boundary propagators, which are the solutions to the equations for the $A_M$ and $V_M$ fields, also will coincide at this approximation. In the finite temperature case, in the soft-wall model with the thermal dilaton, we use formal replacements $z\rightarrow r, k\rightarrow K(T) $, which do not change the EOMs. Consequently, the known expression for the vector bulk-to-boundary propagator at finite temperature obtained in Ref. \cite{Gutsche2} can be used for the axial-vector field's propagator as well. In the soft-wall model with the thermal dilaton field $\varphi (r, T)$ defined in Eq. (\ref{11}), the EOM for a vector field was obtained in the following form  \cite{Gutsche2, Gutsche3}:
	\begin{equation}
		\partial_{r}\left(\frac{e^{-\varphi(r, T)}}{r}\partial_{r}A(Q, r, T)\right)- Q^{2}\frac{e^{-\varphi(r, T)}}{r}\partial_{r}A(Q, r, T)=0.
		\label{14}
	\end{equation}
	The $A(Q,r,T)$ thermal propagator $A(Q,r,T)$ satisfies the boundary condition $A(Q, z=0, T)=1$.
	 The Eq. (\ref{14}) is similar to one at zero temperature, and the only difference between them is the $T$-dependence of the dilaton parameter. 
	So, the solution of the  Eq. (\ref{14}) written as \cite{Gutsche3} 
	\begin{equation}
		A(Q, r, T)=\Gamma \left(1+a(Q, T)\right)U\left(a(Q, T), 0, K^{2}r^{2}\right)=K^{2}r^{2}\int_{0}^{1}\frac{dx}{(1-x)^{2}}x^{a(Q, T)}e^{-K^{2}r^{2}\frac{x}{1-x}},
		\label{15}
	\end{equation}
	is the bulk-to-boundary propagator for the axial-vector field also. Here $\Gamma$ is Euler's gamma function and $a(Q,T)=\frac{Q^{2}}{4K^{2}(T)}$. $U(x,y,z)$ is a Tricomi function, also known as the confluent hypergeometric function of the second kind.
	\section{Holography and the bulk interaction Lagrangian}
	We can apply the holography principle to get the $G_A$ form factor in the boundary QCD from the bulk interaction action:
	\begin{equation}
	S_{int}=\int d^5x \sqrt{g} L_{int}. \label{38}
	\end{equation}
	The generating functional $Z_{AdS}$ of the bulk theory is defined by the classical bulk action $S_{int}$:
	\begin{equation}
	Z_{AdS}=e^{iS_{int}}.
	\label{39}
	\end{equation}
	AdS/CFT correspondence identifies the generating functional $Z_{AdS}$ with the generating function $Z_{QCD}$ of the boundary QCD \cite{Makoto}:
	\begin{equation}
	Z_{AdS}=Z_{QCD}.
	\label{40}
	\end{equation}
	After calculating  $S_{int}$ for the bulk theory and using the holographic identification in Eq. (\ref{40}), one can find the $<J_{\mu}^a>^{QCD}$  axial-vector current of nucleons in the boundary QCD theory. The variation derivative of the gravity functional $Z_{AdS}$ over the $A_{\mu}^{0} $  field will give us sought current:
	\begin{equation}
	<J_{\mu}^a>^{QCD}=-i\frac{\delta Z_{AdS}}{\delta A_{\mu}^{a (0)}}|_{A_{\mu}^{a (0)}=0}.
	\label{41}
	\end{equation}
	Here $A_{\mu}^{(0)} $ is the UV boundary value of the $A_{\mu}$ bulk axial-vector field and has an interpretation of a wave function of the axial-vector meson at the boundary.  
	The axial-vector current $J_{\mu}^a(p^{\prime},p)=G_A\bar{u}(p^{\prime})\gamma^5\gamma_{\mu}\left(\tau^a/2\right)u(p)$ obtained from the (\ref{41})  contains the $G_A$ factor, which is the integral over the $r$ coordinate and depends on $Q^2$. According to AdS/CFT correspondence the $\bar{u}(p^{\prime})\gamma^5\gamma_{\mu}\left(\tau^a/2\right)u(p)$ current is the axial-vector current of the nucleons. Consequently, the $G_A\left(Q^2 \right)$ factor is accepted as the axial-vector form factor of the nucleons. This correspondence is relevant for the bulk theory at finite temperature as well.
	
	Now, the main question is how to determine an explicit form of the $L_{int}$ interaction Lagrangian between the $A_M$, $X$, and $\Psi_{1,2}$ fields in the bulk theory. On the one hand, it is possible to construct different Lagrangian terms, which describe the different interactions between these fields. However, on the other hand, we are interested in the interaction terms, which produce namely the axial-vector current in Eq. (\ref{1}) in the boundary theory. These terms should be proportional to the $A_M$ axial-vector field because of the $P$ invariancy of the Lagrangian and be 5D Lorentz invariant. The necessary terms can be taken from the zero-temperature case and then extended to the finite-temperature case by replacing the fields at $T=0$ with the thermal ones. The interactions between the bulk fields $A_{M}$, $X$, and $\Psi_{1,2} $ at zero-temperature case presented in the Refs.  \cite{Chen,Mamedov,Gutsche1,Atayev1,Atayev2,Huseynova2}, will contribute to the finite-temperature $G_{A}\left( Q^{2}, T\right) $ form factor as well. These terms can be classified as follows:
 
	1)  a minimal coupling term
	\begin{equation}
		\textit{L}^{(1)}=\overline{\Psi }_{1}\Gamma ^{M}\left(A_{L}\right)_M\Psi _{1}-\overline{\Psi }_{2}\Gamma ^{M}\left(A_{R}\right)_M\Psi _{2}=\frac{1}{2}\left(\overline{\Psi }_{1}\Gamma ^{M}A_{M}\Psi _{1}-\overline{\Psi }_{2}\Gamma ^{M}A_{M}\Psi_{2}\right),
		\label{42}
	\end{equation}
	2) a magnetic gauge coupling term
	\begin{eqnarray}
		\textit{L}^{(2)}=ik_1\left\{\overline{\Psi }_{1}\Gamma^{MN}\left(F_L\right)_{MN}\Psi _{1}-\overline{\Psi}_{2}\Gamma^{MN} \left(F_R\right)_{MN}\Psi_{2}\right\}\nonumber\\ =\frac{i}{2}k_1\left\{\overline{\Psi }_{1}\Gamma^{MN}F_{MN}\Psi _{1}+\overline{\Psi }_{2}\Gamma^{MN}F_{MN}\Psi_{2}\right\},
		\label{43}
	\end{eqnarray}
	where $F_{MN}=\partial_MA_N-\partial_NA_M$ is the field stress tensor of the axial-vector field $A_M$.
	
	3)  Three-field interaction term, which was introduced in Ref. \cite{Mamedov}:
	\begin{eqnarray}
		\textit{L}^{(3)}=\frac{g_Y}{2}\left[\overline{\Psi }_{1}X \Gamma
		^{M}\left(A_{L}\right)_M\Psi_{2}-\overline{\Psi }_{2}X^{\dagger}
		\Gamma ^{M}\left(A_{R}\right)_M\Psi_{1}+h.c.\right] \nonumber \\
		=g_Y\left(\overline{\Psi }_{1}X\Gamma^{M}A_{M}\Psi_{2}+\overline{\Psi }_{2}X^{\dagger}\Gamma ^{M}A_{M}\Psi_{1}\right).
		\label{44}
	\end{eqnarray}
	The last term describes an interaction of three bulk fields at one point, and corresponds to the nucleon-double meson interaction at the boundary. Since this interaction is the chirality-changing one, the Yukawa coupling constant $g_Y$ was chosen as the coupling constant in this term. Direct calculations show that $\textit{L}^{(3)}$ term contributes to the axial-vector form factor of nucleons.

	\section{$G_A$ form factor at finite temperature}
	We write the action terms in the momentum space, using the 
	\begin{equation}
		j^{5\mu}\left(p^{\prime},p \right)=\overline{u}\left( p^{\prime }\right) \gamma
		^{5}\gamma ^{\mu }\frac{\tau ^{a}}{2}u\left( p\right)
		\label{45}
	\end{equation}
	short notation for the axial-vector current. Let's  itemize the $S^{(i)}$ action terms corresponding to the $L^{(i)}$ Lagrangians:
	\begin{eqnarray}
		& 1)\ S^{(1)}=\frac{1}{2}\int d^{4}x\ \int_{0}^{\infty}dr\ \sqrt{g}\left\{\overline{\Psi }_{1}\Gamma ^{\mu }A_{\mu }\Psi _{1}-\overline{\Psi }_{2}\Gamma ^{\mu
		}A_{\mu }\Psi _{2}\right\}\nonumber \\
		&=\frac{1}{2}\int d^{4}pd^{4}p^{\prime } j^{5\mu}\left(p^{\prime},p \right)A_{\mu }^{a}\left( Q\right) \int_{0}^{\infty}dr\ A\left( Q, r, T\right) \left[ \left\vert F_{1R}\left( r, T\right) \right\vert
		^{2}-\left\vert F_{1L}\left( r, T\right) \right\vert ^{2}\right],
  \label{46}
	\end{eqnarray}
	\begin{eqnarray}
		& 2)\ S^{(2)}=\frac{i}{4}k_1\int d^{4}x\ \int_{0}^{\infty}dr\ \sqrt{g} \left\{\overline{\Psi }_{1}\left[ \Gamma ^{5},\Gamma ^{\mu}\right]\partial_{5}A_{\mu}\Psi _{1}+ \overline{\Psi }_{2}\left[ \Gamma ^{5},\Gamma ^{\mu}\right]\partial_{5}A_{\mu}\Psi _{2}\right\}\nonumber \\
		&=\frac{k_1}{2} \int d^{4}pd^{4}p^{\prime }j^{5\mu}\left(p^{\prime},p \right)A_{\mu }^{a}\left( Q\right)\int_{0}^{\infty}dr\
		r\left( \partial _{r}A\left( Q,r,T\right) \right) \left[ \left\vert
		F_{1R}\left( r, T\right) \right\vert ^{2}+\left\vert F_{1L}\left( r, T\right)
		\right\vert^{2}\right],
		\label{47}
	\end{eqnarray}
	\begin{eqnarray}
		& 3)\ S^{(3)}=g_Y\int d^{4}x\ \int_{0}^{\infty}dr\ \sqrt{g}\left\{\overline{\Psi }_{1}X\Gamma ^{\mu }A_{\mu }\Psi _{2}+\overline{\Psi }_{2}X^{\dagger}\Gamma^{\mu}A_{\mu }\Psi _{1}\right\}\nonumber \\
		&=g_Y\int d^{4}pd^{4}p^{\prime }j^{5\mu}\left(p^{\prime},p \right)A_{\mu }^{a}\left( Q\right)\int_{0}^{\infty}dr A\left( Q,r, T\right) 2 v\left(r, T\right) F_{1L}\left( r, T\right) F_{1R}\left( r, T\right).
		\label{48}
	\end{eqnarray}
	Here the both ($\overline{\Psi }$ and $\Psi$) bulk fermions are considered on the mass shell  $\left(|p|=|p^{\prime}|=m\right)$, and $m$ is the mass of the nucleons. 
	Total action $S_{int}=S^{(1)}+S^{(2)}+S^{(3)}$ will produce the axial-vector form factor  $G_A$ of the nucleons at finite temperature.
	Taking derivatives over $A_{\mu }^{a}\left( Q \right) $ from the $S^{(i)}$ action terms we get the $G_A^{(i)}(T)$ contributions of these terms into the axial-vector form factor $G_{A}\left(Q^{2},T\right)$:
	\begin{equation}
		1) \ G^{(1)}_{A}\left( Q^{2}, T\right) =\frac{1}{2}\int_{0}^{\infty}dr A\left(Q, r, T\right) \left[ \left\vert F_{1R}\left( r,T\right) \right\vert^{2}-\left\vert F_{1L}\left(r,T\right) \right\vert^{2}\right],
		\label{49}
	\end{equation}
	\begin{equation}
		2) \ G^{(2)}_{A}\left(Q^{2}, T\right) =\frac{k_1}{2}\int_{0}^{\infty}drr\left(\partial_{r}A\left( Q,r, T\right) \right) \left[ \left\vert F_{1R}\left(r, T\right) \right\vert ^{2}+\left\vert F_{1L}\left( r, T\right) \right\vert ^{2}
		\right],
		\label{50}
	\end{equation}
	\begin{equation}
		3) \ G^{(3)}_{A}\left( Q^{2}, T\right)=\frac{1}{2} g_Y\int_{0}^{\infty}dr \
		A\left( Q, r, T\right)  v\left(r, T\right) F_{1L}\left( r, T\right) F_{1R}\left( r, T\right).
		\label{51}
	\end{equation}
	\begin{equation}
		G_{A}\left( Q^{2}, T\right)= G^{(1)}_{A}\left( Q^{2}, T\right)+ G^{(2)}_{A}\left(Q^{2}, T\right)+G^{(3)}_{A}\left( Q^{2}, T\right).
		\label{52}
	\end{equation}
\section{Axial-vector transition form factor}
 A formalism for the study of the nucleon resonances within the holographic model has been proposed in Refs. \cite{Teramond,Gutsche20,Ramalho3,Ramalho4,Gutsche22}.
 In the present work for the transition form factor, we can apply the interaction Lagrangians (\ref{42})-(\ref{44}). However, this time the $\overline{\Psi }_{1,2}$ states describe the excited nucleon states, while $\Psi _{1,2}$ ones are the ground states. So, in the profile functions $ F_{1,2 L}^{(n)*}$ we shall set $n=1$ and for this KK mode we shall take the mass $m^*$ of the excited nucleon $\left(|p^{\prime}|=m^*\right)$.  In the  profile functions $ F_{1,2 L}^{(n)}$ we set $n=0$ and $\left(|p|=m\right)$.
  By getting one of the nucleon profile functions excited state we obtain the $G_{AT}^{(i)}(T)$ contributions of these terms into the axial-vector transition form factor  $G_{AT}\left(Q^{2}, T\right)$:
	\begin{equation}
	\	1) G^{(1)}_{AT}\left( Q^{2}, T\right) =\frac{1}{2}\int_{0}^{\infty}dr A\left(Q, r, T\right) \left[
  \left( F_{1L}^{(n)*}F_{1L}^{(m)}-F_{2L}^{(n)*}F_{2L}^{(m)}\right)\right]
	\label{53}
	\end{equation}
	\begin{equation}
	\	2) \ G^{(2)}_{AT}\left(Q^{2}, T\right) =\frac{k_1}{2}\int_{0}^{\infty}dr\partial_{r}A\left( Q,r, T\right)\left[ 
  \left( F_{1L}^{(n)*}F_{1L}^{(m)}+F_{2L}^{(n)*}F_{2L}^{(m)}\right)\right]
		\label{54}
	\end{equation}
	\begin{equation}
	\	3) \ G^{(3)}_{AT}\left( Q^{2}, T\right)=\frac{1}{4} g_Y\int_{0}^{\infty}dr \
		A\left( Q, r, T\right)  v\left(r,T\right)\left[ 
  \left( F_{1L}^{(n)*}F_{1R}^{(m)}-F_{2L}^{(n)*}F_{2R}^{(m)}\right)\right].
		\label{55}
	\end{equation}
	\begin{equation}
		G_{AT}\left( Q^{2}, T\right)= G^{(1)}_{AT}\left( Q^{2}, T\right)+ G^{(2)}_{AT}\left(Q^{2}, T\right)+G^{(3)}_{AT}\left( Q^{2}, T\right).
		\label{56}
	\end{equation}
	 There are the following relations between the profile functions of the first and second bulk fermion fields:
$F_{1L}=F_{2R},\quad F_{1R}=-F_{2L}.$

	\section{Numerical results}
 
  In Fig. 1, we plot the temperature dependency of nucleon mass. This dependence has a shape similar to one for the meson mass spectrum in Ref. \cite{Gutsche2}.
  
   In Figs. 2-4, we present the dependencies of the normalized $G_A\left(Q^2,T\right)/G_A\left(0,0 \right)$ form factor  on $T$ and $Q^2$ for the ground state $(n=0)$ and the first excited state $(n=1)$ of nucleons  in the left  and right panels correspondingly. To see the impact of the parameter $\alpha$ on the shape of the form factor dependency, we have plotted its graph for the values $\alpha$ = 0.1, 0.2, 0.3. 
    It is seen that the form factor graphs at different values of $\alpha$ have the same shape for the ground and excited states correspondingly. This means that the impact of the quantum number $n$ on the shape of the form factor is weak enough. The shape deformation of the form factor on changing the parameter $\alpha$  may observe by comparing the left panel figures one with another (or right ones). We collect the $G_A\left(Q^2, T \right)/G_A\left(0,0 \right)$ form factor graphs at $Q^2=0$ in Fig.5 to show more apparent the shape deformation on changing the $\alpha$ parameter.  Another observation is that in all  $T$-dependency graphs the $G_A$ form factor goes to zero around the $T_c=0.2$ GeV. This is interpreted with the hadron melting at the temperature of the confinement-deconfinement phase transition. In Fig.6, we plot two-dimensional graphs for the form factor at fixed values $T=0, 0.1, 0.17$ GeV. This analysis shows more clearly how the shape of the $Q^2$-dependency deforms on a small change in $T$.

The main aim of the present work is to study the influence of the medium temperature on the $G_A$ form factor. Since there is no experimental data\footnote{The only theoretical result for the temperature-dependent $g_{a}(T)$ coupling constant is  Ref. \cite{Dominguez, {Dominguez3}},  which is based on 
QCD Sum Rules. In Ref. \cite{Dominguez2} these authors express that this result contrasts with other effective hadronic couplings (\cite{Dominguez2, Ayala}) and diverges when $T\rightarrow T_{c}$. This result also disagrees with the soft-wall model result on coupling (form factor) vanishing at $T_c$ observed in \cite{Gutsche2, Gutsche3, Gutsche4, Nasibova1, Taghiyeva}.} for the  $G_A$ form factor at finite temperature, in Fig. 7, we present a comparison of our result for $G_{A}(Q^2)/G_{A}(0)$ at $T=0$ with the hard-wall AdS\ QCD model result for the $G_{A}(Q^2)$ form factor \cite{Mamedov, Huseynova, Huseynov}, experimental data \cite{Bernard,Park} and the result of the Light Cone Sum Rules (LCSR) \cite{Anikin}.

  All theoretical predictions for the $Q^2$ dependencies have the form   $1/Q^n$ dependence, which is typical for the nucleon form factors.  Experimental data for $G_{A}$ are exist in two different intervals of $Q^2$, in  $~0.075$ GeV$^2$ $\leq$ $Q^2$  $\leq1$ GeV$^2$  (\cite{Bernard}) and in $2$ GeV$^2$ $\leq Q^{2}$ $\leq 4$ $GeV^2$ (\cite{Park}).

  There is no momentum restriction on the applicability of holographic models, while other theoretical approaches, such as light cone sum rules, are applicable only for $Q^2 \ge 1$ GeV$^2$. The theoretical dependencies obtained from the hard- and soft-wall models cover both momentum intervals of the experiment.

	  It is seen from Fig.7 that holographic model predictions for the $Q^2$ dependence of the $G_A$ form factor are in good agreement with the experimental data and other theoretical approaches.

	  In numerical calculations, we have used for the values of light quark mass $m_{q}$ and quark condensate $\Sigma$ the values $m_q=0.00234$ GeV and $\left(\Sigma\right)^{1/3}=0.311$ GeV respectively, \cite{Maru}. The constant $k_{1}=-0.98$ was taken from  \cite{Ahn}, which was obtained from the fitting of the couplings $g_{\pi NN}$ and $g_{\rho NN}$ of  AdS/QCD model with the experimental data. The Yukawa constant $g_Y=9.182$ is fixed as in \cite{Hong}.
   In Fig.8, we plot the axial-vector transition form factor $G_{AT}$ for the above-fixed values of the $\alpha$ parameter using formulas (\ref{49})-(\ref{52}). In Fig.9, similar analysis for the shape deformation at temperature changes has been carried out.
	  In comparison with the $G_A$ form factor, we observe some small changes in the shape of the graphs for the $G_{AT}$ form factor though forms of all dependencies in $G_{AT}$ form factor are similar to the ones in corresponding $G_A$ form factor.

	\section{Summary}
	 In the present work, we have investigated the temperature dependence of the axial-vector form factor in the framework of the soft-wall AdS/QCD model containing the thermal dilaton field. Numerical analysis shows that the form factor decreases with increasing temperature. This means in a hot medium, the $\beta$   decay has less  probability when temperature increases.  This result may be used for the studies of nucleon-axial-vector meson interactions and the influence of temperature on $beta$ decay in the nuclear

medium.

 \begin{figure}[!ht]
		\centering
		\includegraphics[scale=0.7]{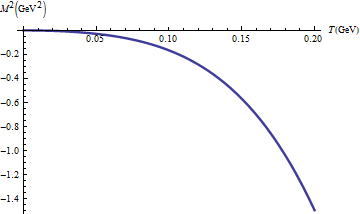}
		\caption{$T$ dependence of nucleon mass $M_{0,\frac{1}{2}}^{2}(T)$.}
	\end{figure}
	
	\begin{figure}[!ht]
		\centering
		\includegraphics[scale=0.5]{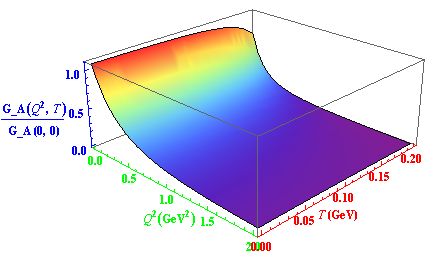}
		\includegraphics[scale=0.5]{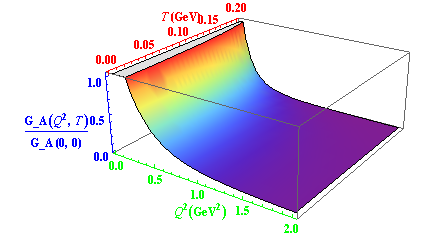}
		\caption{$G_{A}(Q^2,T)/G_{A}(0,0)$ form factor in the ground $(n=0)$ and excited states $(n=1)$ of the nucleons at $\alpha=0.1$.}
	\end{figure}
	\begin{figure}[!ht]
		\centering
		\includegraphics[scale=0.5]{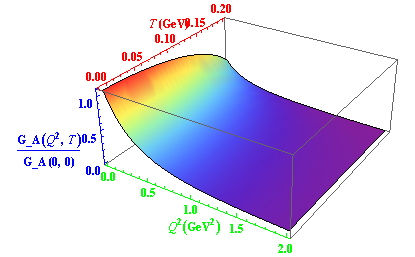}
		\includegraphics[scale=0.5]{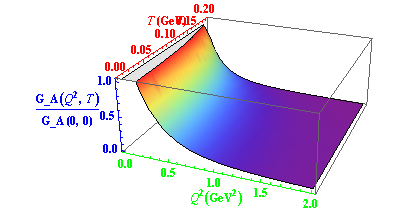}
		\caption{$G_{A}(Q^2,T)/G_{A}(0,0)$ form factor in the ground $(n=0)$ and excited states $(n=1)$ of the nucleons at $\alpha=0.2$.}
	\end{figure}
	\begin{figure}[!ht]
		\centering
		\label{fig:Figure1}
		\includegraphics[scale=0.5]{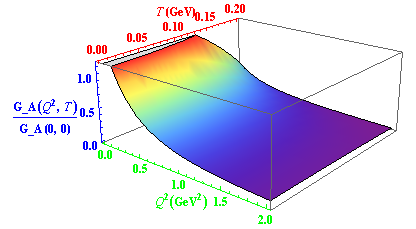}
		\includegraphics[scale=0.5]{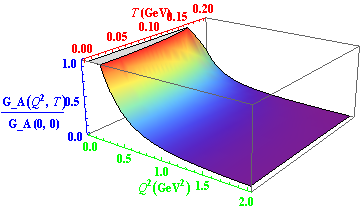}
	\caption{$G_{A}(Q^2,T)/G_{A}(0,0)$ form factor in the ground $(n=0)$ and excited states $(n=1)$ of the nucleons at $\alpha=0.3$.}
	\end{figure}
	\begin{figure}[!ht]
		\centering
		\includegraphics[scale=0.5]{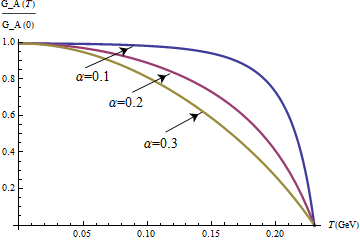}
		\includegraphics[scale=0.5]{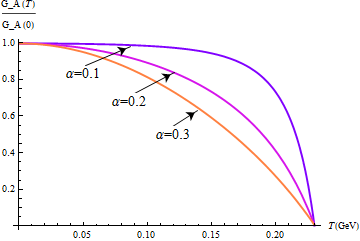}
		\caption{ $G_{A}(T)/G_{A}(0)$ form factors at $Q^2=0$ and different values of $\alpha$ in the ground $(n=0)$ and excited $(n=1)$ states of the nucleons.}
	\end{figure}
	\begin{figure}[!ht]
	\centering
	\label{fig: Figure 5}
	\includegraphics[scale=0.5]{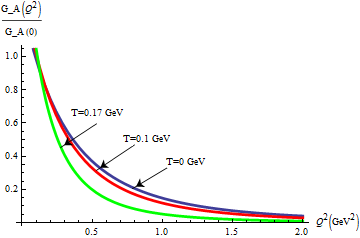}
	\includegraphics[scale=0.5]{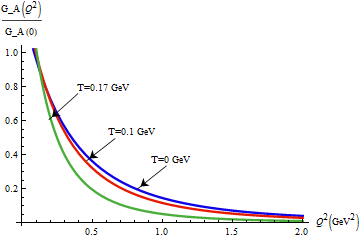}
	\caption{ $G_{A}(Q^2)/G_{A}(0)$ form factors at different values of $T$ in the ground $(n=0)$ and excited $(n=1)$ states of the nucleons.}
	\end{figure}
\begin{figure}[!ht]
		\includegraphics[scale=0.7]{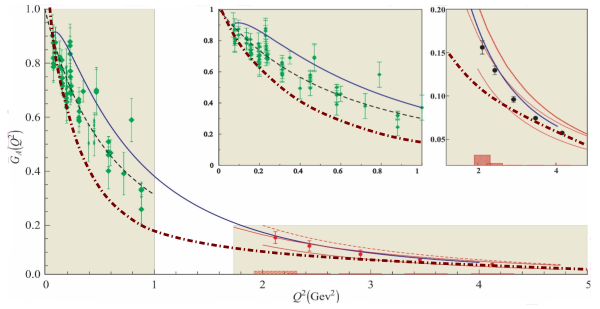}
	\caption{ Comparison of our result for the $G_{A}(Q^2, T=0)/G_{A}(0,0)$ form factor at $n=0$ and $T=0$ values (densely dash-dotted line) with the hard-wall result obtained in \cite{Mamedov,Park} (solid blue line), LCSR results in the $2\leq Q^2\leq 5$ interval (red dashed line and red solid line) (\cite{Anikin}) and experimental data  taken from \cite{Bernard} and \cite{Dominguez2}.} 
\end{figure}
\begin{figure}[!ht]
\centering
\includegraphics[scale=0.5]{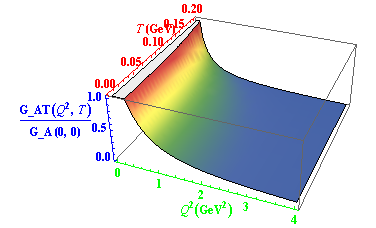}
\includegraphics[scale=0.5]{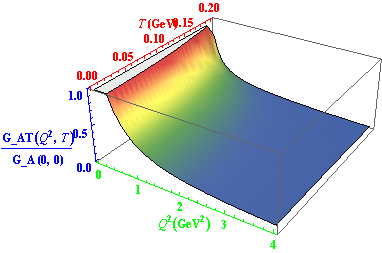}
\includegraphics[scale=0.5]{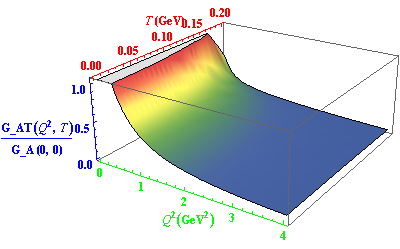}
\caption{$G_{AT}(Q^2,T)/G_{AT}(0,0)$ transition form factor of  nucleons at parameter $\alpha=0.1, 0.2, 0.3$ respectively}
\end{figure}
	\begin{figure}[!ht]
		\centering
		\includegraphics[scale=0.5]{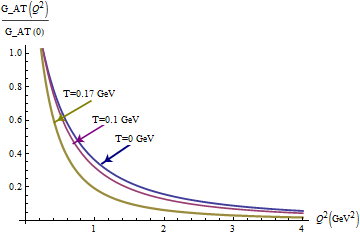}
		\includegraphics[scale=0.5]{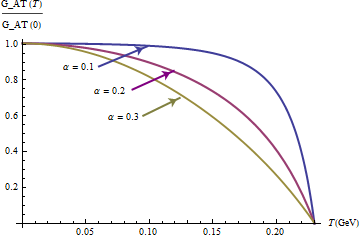}
		\caption{$Q^2$ dependence of  transition form factor  $G_{AT}(Q^2)/G_{AT}(0)$ at different values of $T$ and  $T$ dependence of $G_{AT}(T)/G_{AT}(0)$  at different values of parameter $\alpha$  respectively}
	\end{figure}
 
\end{document}